\documentclass[%
 reprint,
 amsmath,amssymb,
 aps,
 prl,
]{revtex4-2}

\usepackage{graphicx}
\usepackage{dcolumn}
\usepackage{bm}

\usepackage[capitalise]{cleveref}

\begin{document}

\title{The complete control of scattering waves in multi-channel structures}

\author{Qi Gao}
\author{Yun-Song Zhou}
 \email{zyspaper@cnu.edu.cn}
\author{Li-Ming Zhao}
\affiliation{
 Department of Physics, Capital Normal University, Beijing, 100048. China
}
\date{\today}

\begin{abstract}
The issue of photon spin Hall effect was generalized as a universal question of how to control all the scattering waves in a multi-channel structure (complete control). A general theory was proposed, which provides a simple way to achieve the complete control. This theory shows also that the necessary condition for complete control is that the structure must contain a complete set of sources. To demonstrate the application of the theory, the typical scattering patterns in the two-channel and four-channel structures are achieved theoretically.   Previous this research, one could only artificially control the scattering waves in two channels out of a four-channel structure.
\end{abstract}

\maketitle

The new kind of photonic spin Hall effect (PHE) attracted great interest\cite{lee2012role,rodriguez2013near,rodriguez2014sorting,rodriguez2014resolving,mitsch2014quantum,petersen2014chiral,kapitanova2014photonic,espinosa2016transverse,coles2016chirality,picardi2017unidirectional} because of its application prospects and interesting theoretical problems\cite{bliokh2014extraordinary,bliokh2015transverse,bliokh2015spin,bliokh2015quantum,van2016universal,picardi2018janus,gao2021continuous}. Considering this problem from a general perspective, it belongs to the scattering control problem. Taking the PHE on metal surface\cite{lee2012role,rodriguez2013near} as an example, it is controlled by the polarization state of incident light. The surface polariton plasmon (SPP) wave can be controlled traveling to left (left channel) or right (right channel) artificially, here a channel corresponds to a guide mode. So, the metal surface is a two-channel structure, the two channels are separated at the point where the scatterer is located. Most of the works currently are focused on the two-channel structures, but some researches are also extended to the four-channel structures\cite{picardi2018janus,gao2021continuous,luxmoore2013interfacing,wang2017photonic,zhao2021tunable}. However, if we review these works on four-channel structures, a common point can be found that, only two channels can be selected from the four channels, in which scattering waves are controllable as people whish. The other waves in the remaining two channels can not be controlled artificially (known later that the waves in remaining channels are determined by the waves in the selected channels). Of course, the two controllable channels are optional. This kind of control can be name as 2/4 control. (noticing that the PHE on metal surface is 2/2 control because the waves in the two channels can be distributed with any proportion). Thus, a series of interesting questions arose: how to realize 3/4, and 4/4 controls in a four-channel structure? The more universal questions are proposed: in a n-channel structure (if a channel involves two modes it should be regard as two channels), how to realize the complete control(the waves in all channels are controllable)? and what's the condition for achieving the complete control?

\begin{figure}[t]
\includegraphics[width=0.75\columnwidth]{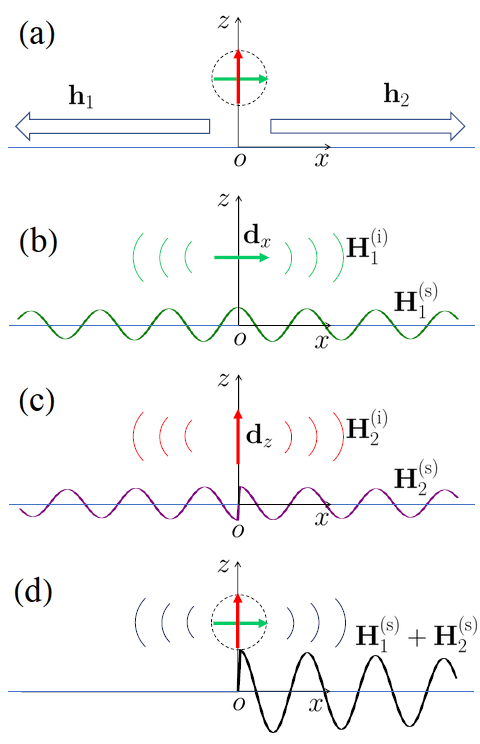}
\caption{\label{fig:1}The photon spin Hall effect on metal surface can be thought as the scattering control process in two-channel structure. (a) the dashed circle denotes the rotational dipole, which can be separated with both the level and vertical dipoles. The surface on the left (right) side of the dipole is channel 1 (channel 2), in which the SPP mode is $\mathbf{h}_1$ ($\mathbf{h}_2$); (b) the SPP wave $\mathbf{H}_1^{({\text{s}})}$ excited by the level dipole. (c) the SPP wave $\mathbf{H}_2^{({\text{s}})}$ excited by the vertical dipole. (d) the total wave $\mathbf{H}^{({\text{s}})}=\mathbf{H}_1^{({\text{s}})}+\mathbf{H}_2^{({\text{s}})}$ scattered by the rotational dipole.}
\end{figure}

To answer these questions, let's first consider the simplest structure\cite{lee2012role,rodriguez2013near}. A notch cut on a metal surface which will divide the surface into two channels, channel 1 on the left side and channel 2 on the right side of the notch. When the notch is illuminated by an incident light, two SPP waves will be excited and transmit respectively in both channels\cite{rodriguez2013near}. The magnetic fields of the eigenmodes in channel 1 and channel 2 are expressed with $\mathbf{h}_1$ and $\mathbf{h}_2$ respectively(see \cref{fig:1}(a)). When the incident light is circularly polarized, the illuminated notch can be thought as a rotational dipole\cite{lee2012role,rodriguez2013near,picardi2018janus,gao2021continuous}, which is composed of two independent dipoles along two perpendicular directions, just as shown in \cref{fig:1}(a). Each independent dipole can be called as a light source. The contributions from the two sources can be discussed independently.

For the $\mathbf{d}_x$ dipole (source 1), its radiation magnetic field is $\mathbf{H}^{(\text{i})}_1$, which will excite $\mathbf{H}^{(\text{s})}_1$, the magnetic field of SPP wave (\cref{fig:1}(b)). $\mathbf{H}^{(\text{s})}_1$ is the total scattering wave involving the both waves in the different directions. In other words $\mathbf{H}^{(\text{s})}_1=s_{11}\mathbf{h}_1+s_{21}\mathbf{h}_2$. As the same reason, the $\mathbf{d}_z$ dipole (source 2) has the relationship $\mathbf{H}^{(\text{s})}_2=s_{12}\mathbf{h}_1+s_{22}\mathbf{h}_2$ (\cref{fig:1}(c)). Thus, we have the matrix equation
\begin{equation}\label{Eq:H=Sh}
\begin{aligned}
\begin{pmatrix}
  {{\mathbf{H}}_1^{({\text{s}})}}&{{\mathbf{H}}_2^{({\text{s}})}}
\end{pmatrix}
 =
\begin{pmatrix}
  {{\mathbf{h}_1}}&{{\mathbf{h}_2}}
\end{pmatrix}
\begin{pmatrix}
  {{s_{11}}}&{{s_{12}}} \\
  {{s_{21}}}&{{s_{22}}}
\end{pmatrix}
\equiv
\begin{pmatrix}
  {{\mathbf{h}_1}}&{{\mathbf{h}_2}}
\end{pmatrix}S
\end{aligned}.
\end{equation}
Here $S$ is named as transform matrix. If each source strength $\mathbf{H}^{(\text{i})}_{n}$ is normalized, or standard with a fixed value, each element of the transform matrix can be determined by theoretical calculation, numerical simulation, or practice measurement, and then the total field excited by the two dipoles should be composed as $\mathbf{H}^{(\text{s})}=\mathbf{H}^{(\text{s})}_1+\mathbf{H}^{(\text{s})}_2$ (sketched in \cref{fig:1}(d)).

However, for a certain structure, the matrix $S$ should be determined first by the normalized or standard $\mathbf{H}^{(\text{i})}_{n}$. In actual problems, the source strength (depend on the incident light) may not be normalized or standard, but $A_1\mathbf{H}^{(\text{i})}_1$ and $A_1\mathbf{H}^{(\text{i})}_2$, the total scattering field should be $\mathbf{H}^{(\text{s})}=A_1\mathbf{H}^{(\text{s})}_1+A_2\mathbf{H}^{(\text{s})}_2$. Combined with \cref{Eq:H=Sh}, we get
\begin{equation}\label{Eq:H=Ch}
{{\mathbf{H}}^{({\text{s}})}} =
\begin{pmatrix}
  {\mathbf{h}_1^{}}&{\mathbf{h}_2^{}}
\end{pmatrix}
\begin{pmatrix}
  {{C_1}} \\
  {{C_2}}
\end{pmatrix},
\end{equation}
where
\begin{equation}\label{Eq:C=SA}
\begin{pmatrix}
  {{C_1}} \\
  {{C_2}}
\end{pmatrix}= S
\begin{pmatrix}
  {{A_1}} \\
  {{A_2}}
\end{pmatrix}.
\end{equation}
\Cref{Eq:C=SA} is a linear transform. If $S$ is an invertible matrix, then
\begin{equation}\label{Eq:A=SC}
\begin{pmatrix}
  {{A_1}} \\
  {{A_2}}
\end{pmatrix} = {S^{ - 1}}
\begin{pmatrix}
  {{C_1}} \\
  {{C_2}}
\end{pmatrix},
\end{equation}
\cref{Eq:H=Ch,Eq:A=SC} provide the way to artificially control the scattering field. For instance, if we hope the scattered light transmitting only in $x$ direction (\cref{fig:1}(d)), the wavefunction should be $\mathbf{H}^{(\text{s})}=C_1\mathbf{h}_1+C_2\mathbf{h}_2$, it means $C_1=0$ and $C_2=1$. The corresponding source strengths $A_1,A_2$ can be obtained from Eq. (4). In general case, any wanted scattering field, which is represented by $C_1,\ C_2$ can be obtained when we set the source strength $A_1,\ A_2$ according to \cref{Eq:A=SC}. Thus, the complete control is achieved. In other words, the waves in two channels can be controlled in the two-channel structure, it is named as 2/2 control.

Now, it's easy to extend this theory to the multi-channel structure with the same way. For a $n$-channel structure, its $n$ eigenfunctions are $\mathbf{h}_1,\dots,\mathbf{h}_n$.  If we set $n$ independent light sources in the structure, the normalized or standard fields are expressed as $\mathbf{H}^{(\text{i})}_1,\dots,\mathbf{H}^{(\text{i})}_n$, which will excite the scattering fields $\mathbf{H}^{(\text{s})}_1,\dots,\mathbf{H}^{(\text{s})}_n$.
When the real radiation fields are
$A_1\mathbf{H}^{(\text{i})}_1,\dots,A_n\mathbf{H}^{(\text{i})}_n$,  the total scattering field will be ${{\mathbf{H}}^{({\text{s}})}} = \sum\limits_{\beta  = 1}^n {{\mathbf{H}}_\beta ^{({\text{s}})}{A_\beta }}  = \sum\limits_{\alpha ,\beta  = 1}^n {{{\mathbf{h}}_\alpha }{S_{\alpha \beta }}{A_\beta }}= \sum\limits_{\alpha= 1}^n {{{\mathbf{h}}_\alpha }{C_{\alpha}}}$.
Here
\begin{equation}\label{Eq:Cn=SA_n}
\begin{pmatrix}
  {{C_1}} \\
   \vdots  \\
  {{C_n}}
\end{pmatrix}
=
\begin{pmatrix}
  {{s_{11}}}& \cdots &{{s_{1n}}} \\
   \vdots & \ddots & \vdots  \\
  {{s_{n1}}}& \cdots &{{s_{nn}}}
\end{pmatrix}
\begin{pmatrix}
  {{A_1}} \\
   \vdots  \\
  {{A_n}}
\end{pmatrix}.
\end{equation}
If the transform matrix $S$ is invertible, the inverted transformation is
\begin{equation}\label{Eq:An=SCn}
\begin{pmatrix}
  {{A_1}} \\
   \vdots  \\
  {{A_n}}
\end{pmatrix}= {S^{ - 1}}
\begin{pmatrix}
  {{C_1}} \\
   \vdots  \\
  {{C_n}}
\end{pmatrix}.
\end{equation}

As the number $C_i$ is the scattering amplitude in channel $i$ and $A_{j}$ is the radiation amplitude from source $j$, all the scattering fields can be controlled artificially according to \cref{Eq:An=SCn}.

There is also a remaining theoretical question to explain. What is the reversible condition for the transform matrix? According to the linear algebra theory, if $\{\mathbf{h}_1,\dots,\mathbf{h}_n\}$ and $\{\mathbf{H}^{({\text{s}})}_1,\dots,\mathbf{H}^{({\text{s}})}_n\}$ are respectively two set of linear independent functions, then $S$ is invertible. First, the channel modes $\{\mathbf{h}_1,\dots,\mathbf{h}_n\}$ are solved from Helmholtz equations with different wavevectors, they of course are linear independent each other. Second, we demonstrate with some examples that $\{\mathbf{H}^{({\text{s}})}_1,\dots,\mathbf{H}^{({\text{s}})}_n\}$  are independent of each other when the sources are set correctly.

Example 1, two dipoles perpendicular to each other on metal surface as shown is \cref{fig:1} (b) and (c), where $\mathbf{H}^{({\text{s}})}_1\ne c\mathbf{H}^{({\text{s}})}_2$, $c$ is an arbitrary constant. In addition to the above perpendicular electric dipoles, the mutually perpendicular magnetic dipoles (or electric quadrupoles), etc., as the sources can excite independent fields. In practice, although a scatterer illuminated by incident light contains many sources, but it is difficult to independently control these sources. Nevertheless, the same type of sources set at different positions can make $\mathbf{H}^{({\text{s}})}_1\ne c\mathbf{H}^{({\text{s}})}_2$. The following case is an example.

\begin{figure}[th]
\includegraphics[width=\columnwidth]{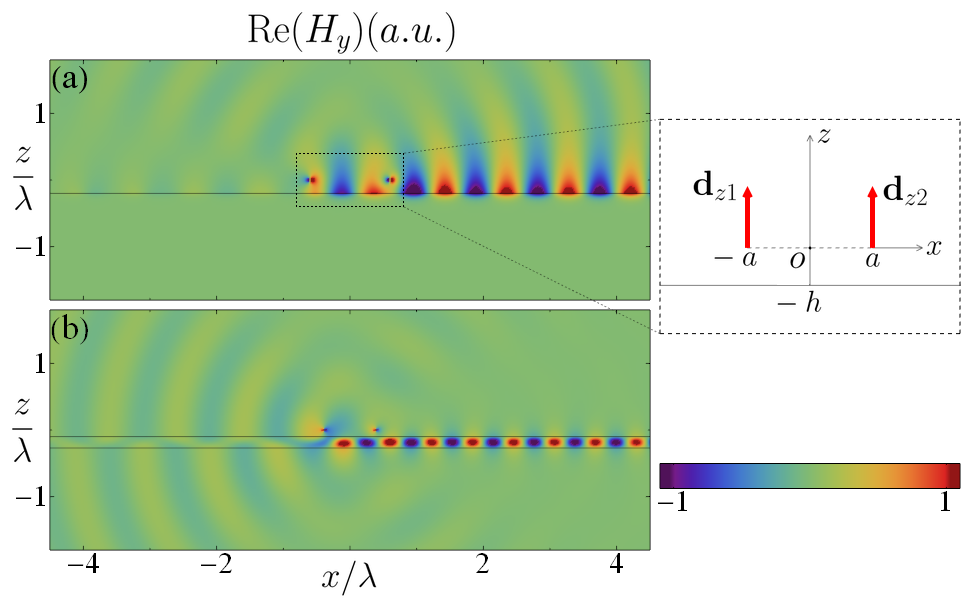}
\caption{\label{fig:2}The unidirectional scattering patterns excited by two vertical dipoles in 2/2 structures. (a) for metal surface system. (b) for dielectric slab system.}
\end{figure}

Example 2, two same electric dipoles $\mathbf{d}_{z1}$ and $\mathbf{d}_{z2}$ with different initial phase are set on the metal surface at different points, just as sketched in \cref{fig:2}. They also have the equation $\mathbf{H}^{{\text{s}}}_1\ne c\mathbf{H}^{{\text{s}}}_2$.

However, the key step to achieve the complete control in $n$-channel structure is to set the $n$ sources correctly to excite $n$ independent fields. In this case, the set of $n$ sources is referred to as a complete set of sources.
In other words, the condition of complete control is a complete set of sources.

In practice, the condition of $|S|\ne0$ can guarantee the $n$ sources form a complete set. If $|S|=0$, these sources should be rearranged. The mathematical expression of the complete set of sources is
\begin{equation}\label{Eq:rankS}
\operatorname{rank}S=n.
\end{equation}

The above discussion is about the case of n sources set in n-channel structure, which is referred to as $n/n$ structure. If $m$ sources are set in a $n$-channel structure ($m/n$ structure), what should happen?

If $m>n$, \cref{Eq:Cn=SA_n} acts as $n$ equations, with $m$ elements $\{A_i\}$. $S$ is a $n\times m$ matrix, if $\operatorname{rank}S=n$, the solutions $\{A_i\}$ always existed regardless of whatever values the $\{C_i\}$ take. The complete control can be achieved. Here $\operatorname{rank}S=n$ means that, there are $n$ sources out of m sources formed a complete set. So the complete control in $m/n$ structure is also expressed with \cref{Eq:rankS}.

When $m<n$, We can select $m$ equations from \cref{Eq:Cn=SA_n}, its coefficient matrix $S_m$ is a $m\times m$ order submatrix. There are two cases are discussed below.

If $|S_m|\ne0$, $S_m$ is an invertible matrix, the corresponding $C_i$ (named selected $C_i$) contained in the selected equations can be assigned with any values, and $A_1,\dots,A_m$ are solved from these equations. The other $C_j$ must be determined by the remaining equations of \cref{Eq:Cn=SA_n}. It means that, the waves in the $m$ channels can be controlled, this is the so-called $m/n$ control, which needs $\operatorname{rank}S=m$.

If $|S_m|=0$, it can not guarantee the selected equations have solutions in the case of that the selected $m$ $C_i$ take arbitrary values. So the $m/n$ control fails.

The both cases of $|S_m|=0$  and $|S_m|\ne0$ may occur in a same $m/n$ structure even the rank $S$ equals to $m$, it depends on the selection of equations. Therefore, in practice, one should first select the wanted channels in which the waves are controllable, and then to set the $m$ sources correctly to insure them independent. In this case the $m/n$ control can be achieved. So in a $m/n$ structure, the most optimal control is $m/n$ control, If the source setting is not appropriate, the $m/n$ control is not available.

Therefore, we can understand these works about four-channel structures\cite{picardi2018janus,gao2021continuous,luxmoore2013interfacing,wang2017photonic,zhao2021tunable}. As the circular polarized incident light contains only two sources, only two channels can be selected as controllable channels. In other words, the most optimal control should be 2/4 control, the 4/4 control is unachievable.

Now we turn to show the applications of this theory in some structures. The most typical PHE is sketched in Fig 1, but we would like to design another novel and simple structure. Assume the same dipoles $\mathbf{d}_{z1}$ and $\mathbf{d}_{z2}$ stand on metal surface at $x=-a$ and $x=a$, respectively (2/2 structure). The channel 1($\mathbf{h}_1$) is in the region of $x<-a$ and channel 2 ($\mathbf{h}_2$) in $x>a$. If the dipole $\mathbf{d}_{z}$ is set at $x=0$, $z=0$ the SPP wave can be written as ${\mathbf{H}^{(\text{s})}} =c\tfrac{x}{|x|}{d_z}{e^{i{k_s}\left| x \right| - \beta z}}\mathbf{\hat{y}}$\cite{gao2021continuous}, here $k_s=k\sqrt{\varepsilon_{\text{metal}}/(\varepsilon_{\text{metal}}+\varepsilon_{\text{air}})}$, $k$ is the wavevector in vacuum (or air), $\beta=\sqrt{k_s^2-k^2}$, $c$ is a constant when the metal, the distance h of the dipole apart from the metal surface, and incident wavelength $\lambda$ are determined. Therefore, when taking $d_{z}=1$ and shifting it to $x=-a$, the $\mathbf{d}_{z1}$ should excites the wave ${\mathbf{H}^{(\text{s})}_1 } =c\tfrac{x+a}{|x+a|}{e^{i{k_s}\left| x+a\right| - \beta z}}\mathbf{\hat{y}}$.
In channel-1, ${\mathbf{H}^{(\text{s})}_1} =-c{e^{i{k_s}\left|x+a\right| - \beta z}}\mathbf{\hat{y}}$, if we define $\mathbf{h}_1=-e^{-ik_x(x+a)-\beta z}\mathbf{\hat{y}}$(amplitude is 1), the element  ${s_{11}} = H_1^{(s)}/{h_1} = c$;
in channel-2, ${\mathbf{H}^{(\text{s})}_1 } =c{e^{i{k_s}(x+a) - \beta  z}}\mathbf{\hat{y}}=ce^{2ik_sa}{e^{i{k_s}(x-a) - \beta z}}\mathbf{\hat{y}}$, $\mathbf{h}_2 = {e^{i{k_s}(x - a) - \beta z}}\mathbf{\hat{y}}$ so ${s_{21}} = c{e^{i2{k_s}a}}$. For the same reason, the dipole $\mathbf{d}_{z2}$ contributes ${s_{12}} = c{e^{i2{k_s}a}}$, ${s_{22}} = c$. Then
\begin{equation}\label{Eq:Sdzdz}
{S^{ - 1}} = \frac{1}{{c(1 - {e^{i2{k_s}a}})}}
\begin{pmatrix}
  1&{ - {e^{i{k_s}a}}} \\
  { - {e^{i{k_s}a}}}&1
\end{pmatrix}.
\end{equation}
If we want the scattering wave to transmit to right, it means $C_1=0$; $C_2=1$, and then \cref{Eq:An=SCn,Eq:Sdzdz} gives
\begin{equation}\label{Eq:A1A2}
{A_1} = \frac{{ - {e^{i{k_s}a}}}}{{c(1 - {e^{i2{k_s}a}})}}, {\text{   }}{A_2} = \frac{1}{{c(1 - {e^{i2{k_s}a}})}}.
\end{equation}

As the above scattering wavefunctions $\mathbf{H}^{(\text{s})}_1$ and $\mathbf{H}^{(\text{s})}_2$ are written in the case of $d_{z1}=d_{z2}=1$. \Cref{Eq:A1A2} means $d_{z1}=A_{1}$ and $d_{z2}=A_{z2}$. In other words, the ratio of ${d_{z2}}/{d_{z1}} = {A_2}/{A_1} =  - {e^{ - i{k_s}a}}$  can cause the scattering to right. For example, when assuming ${\varepsilon _{\text{metal}}} =  - 10 + 0.02i$, $h=0.2\lambda$ and $a=0.6\lambda$, the SPP wavenumber is ${k_s} = (1.0541 + 0.0001171i)k$, then ${d_2}/{d_1} = 0.09347 - 0.99474i$. The field distribution calculated under this condition is displayed in \cref{fig:2}(a), which is consistent with our expectation.

By the way, we consider another 2/2 structure, the metal surface is replaced by a dielectric slab with the thickness of $0.17\lambda $ and dielectric constant of $9 + 0.02i$, $a = 0.39\lambda $, $h=0.1\lambda$ and the waveguide wavenumber is ${k_s} = (1.6108 + 0.003242i)k$. By using the same way, we obtained ${d_2}/{d_1} = 0.03954- 0.98344i$. In this case, the unidirectional scattering result is shown in \cref{fig:2}(b).

\begin{figure}[t]
\includegraphics[width=\columnwidth]{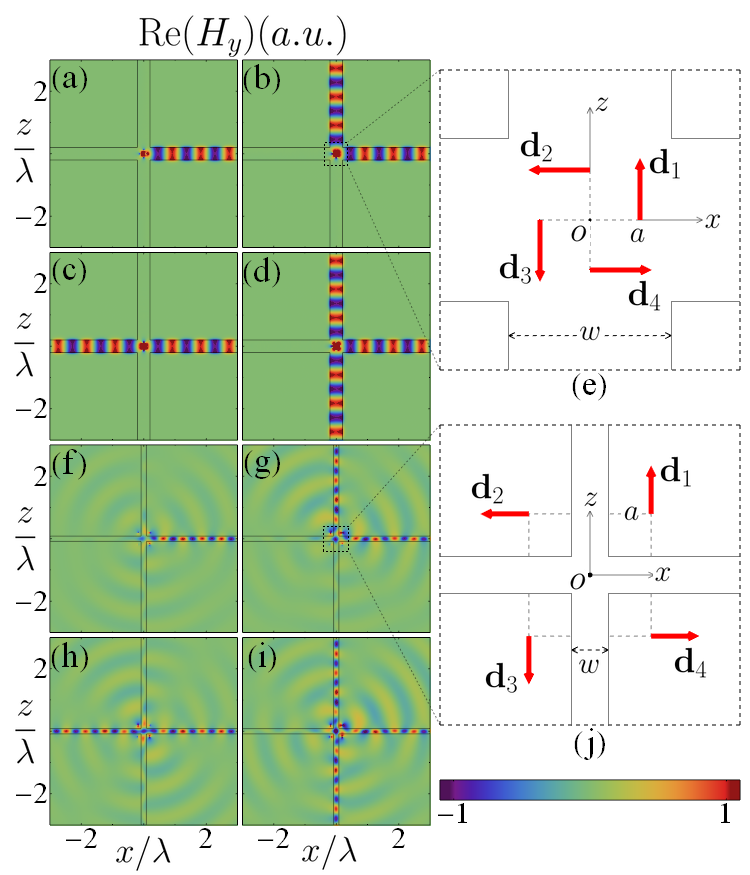}
\caption{\label{fig:4}The four typical scattering patterns in 4/4 structures. (a)-(d) for the structure consisting of two crossed metal slits. The four dipoles are set up as shown in (e). (f)-(i) for the structure composed of two crossed dielectric slabs, the four dipoles are set up as shown in (j).}
\end{figure}

A 4-channel structure can be formed by two crossed metal slits (\cref{fig:4}(a)). The four dipoles $\mathbf{d}_1$, $\mathbf{d}_2$, $\mathbf{d}_3$, and $\mathbf{d}_4$ are set in different positions in different directions, as shown in \cref{fig:4}(e). This creates a 4/4 structure. It's better to get S by numerical simulation method. Since this structure has the $\pi/4$ rotational symmetry, the simulation result from one dipole is sufficient. For instance, excluding the dipoles $\mathbf{d}_2$, $\mathbf{d}_3$, and $\mathbf{d}_4$, only $\mathbf{d}_1$ stands at the original position, the excited SPP waves in channel-1 ($x>0$ direction ), channel-2 ($z>0$ direction), channel-3 ($x<0$ direction), and channel-4 ($z<0$ direction) are denoted with $A$, $B$, $C$, $D$, respectively. Thus $s_{11}=A$, $s_{21}=B$, $s_{31}=C$, and $s_{41}=D$. As the symmetry, the isolated $\mathbf{d}_2$ will contribute the result of $s_{12}=D$, $s_{22}=A$, $s_{32}=B$, and $s_{42}=C$. And so on, the transform matrix is
\begin{equation}\label{Eq:SABCD}
S ={
\begin{pmatrix}
  A&D&C&B \\
  B&A&D&C \\
  C&B&A&D \\
  D&C&B&A
\end{pmatrix}},
\end{equation}
when assuming $\varepsilon_{\text{metal}}=- 20 + 0.02i$ ,$a=0.1\lambda $  and the slit width is $w=0.4\lambda $ . In these slits the SPP wavenumber is ${k_s} = (1.0965 + 0.0000539969i)k$. We can get $A=-0.58683-0.80182i$, $B=D=0.02201 + 0.29451i$ and $C=-0.18300+0.83566i$ then $S$ and $S^{-1}$ are obtained.

For this four-channel structure, there are four typical scattering distribution patterns, they are shown in \cref{fig:4}(a), (b), (c), and (d), and named as a, b, c, and d patterns, respectively. For the a-pattern, $C_1=1$ and $C_2=C_3=C_4=0$, then $A_1,\cdots,A_4$ are obtained from \cref{Eq:An=SCn}. With the results of \cref{Eq:An=SCn}, the actual simulation result is shown in \cref{fig:4}(a). With the same way, the other results are displayed in \cref{fig:4}(b), (c), and (d), respectively.

Another four-channel structure can be composed with two crossed dielectric slabs, the four dipoles are set as shown in \cref{fig:4}(j). Assuming the dielectric constant is  $9 + 0.02i$, $w=0.17\lambda$, and $a=0.1\lambda $. The a, b, c, d patterns are obtained, and put in \cref{fig:4}(f), (g),(h), and (i), respectively.

In summary, the scattering control theory in multi-channel structures are proposed. The complete control requires a complete set of sources. The application of this theory in a $n/n$ structure includes the following four steps: first, determining the eigenmode in each channel; second, determining the scattering waves excited by each source with unit or standard amplitude; third, obtaining the transform matrix; and fourth, acquiring and setting the amplitude of each source though the transform formula (\cref{Eq:An=SCn}) according to our control intention. As the application demonstrations, the complete controls for two-channel structures and four-channel structures are realized theoretically.
For a $m/n$ structure ($m<n$), one should first select the wanted channels and set the sources correctly, then the $m/n$ control can be realized. When $m>n$, the complete control needs at least $n$ sources to form a complete set.


\begin{thebibliography}{20}%
\makeatletter
\providecommand \@ifxundefined [1]{%
 \@ifx{#1\undefined}
}%
\providecommand \@ifnum [1]{%
 \ifnum #1\expandafter \@firstoftwo
 \else \expandafter \@secondoftwo
 \fi
}%
\providecommand \@ifx [1]{%
 \ifx #1\expandafter \@firstoftwo
 \else \expandafter \@secondoftwo
 \fi
}%
\providecommand \natexlab [1]{#1}%
\providecommand \enquote  [1]{``#1''}%
\providecommand \bibnamefont  [1]{#1}%
\providecommand \bibfnamefont [1]{#1}%
\providecommand \citenamefont [1]{#1}%
\providecommand \href@noop [0]{\@secondoftwo}%
\providecommand \href [0]{\begingroup \@sanitize@url \@href}%
\providecommand \@href[1]{\@@startlink{#1}\@@href}%
\providecommand \@@href[1]{\endgroup#1\@@endlink}%
\providecommand \@sanitize@url [0]{\catcode `\\12\catcode `\$12\catcode
  `\&12\catcode `\#12\catcode `\^12\catcode `\_12\catcode `\%12\relax}%
\providecommand \@@startlink[1]{}%
\providecommand \@@endlink[0]{}%
\providecommand \url  [0]{\begingroup\@sanitize@url \@url }%
\providecommand \@url [1]{\endgroup\@href {#1}{\urlprefix }}%
\providecommand \urlprefix  [0]{URL }%
\providecommand \Eprint [0]{\href }%
\providecommand \doibase [0]{https://doi.org/}%
\providecommand \selectlanguage [0]{\@gobble}%
\providecommand \bibinfo  [0]{\@secondoftwo}%
\providecommand \bibfield  [0]{\@secondoftwo}%
\providecommand \translation [1]{[#1]}%
\providecommand \BibitemOpen [0]{}%
\providecommand \bibitemStop [0]{}%
\providecommand \bibitemNoStop [0]{.\EOS\space}%
\providecommand \EOS [0]{\spacefactor3000\relax}%
\providecommand \BibitemShut  [1]{\csname bibitem#1\endcsname}%
\let\auto@bib@innerbib\@empty
\bibitem [{\citenamefont {Lee}\ \emph {et~al.}(2012)\citenamefont {Lee},
  \citenamefont {Lee}, \citenamefont {Park}, \citenamefont {Oh}, \citenamefont
  {Lee}, \citenamefont {Kim},\ and\ \citenamefont {Lee}}]{lee2012role}%
  \BibitemOpen
  \bibfield  {author} {\bibinfo {author} {\bibfnamefont {S.-Y.}\ \bibnamefont
  {Lee}}, \bibinfo {author} {\bibfnamefont {I.-M.}\ \bibnamefont {Lee}},
  \bibinfo {author} {\bibfnamefont {J.}~\bibnamefont {Park}}, \bibinfo {author}
  {\bibfnamefont {S.}~\bibnamefont {Oh}}, \bibinfo {author} {\bibfnamefont
  {W.}~\bibnamefont {Lee}}, \bibinfo {author} {\bibfnamefont {K.-Y.}\
  \bibnamefont {Kim}},\ and\ \bibinfo {author} {\bibfnamefont {B.}~\bibnamefont
  {Lee}},\ }\bibfield  {title} {\bibinfo {title} {Role of magnetic induction
  currents in nanoslit excitation of surface plasmon polaritons},\ }\href@noop
  {} {\bibfield  {journal} {\bibinfo  {journal} {Physical review letters}\
  }\textbf {\bibinfo {volume} {108}},\ \bibinfo {pages} {213907} (\bibinfo
  {year} {2012})}\BibitemShut {NoStop}%
\bibitem [{\citenamefont {Rodr{\'\i}guez-Fortu{\~n}o}\ \emph
  {et~al.}(2013)\citenamefont {Rodr{\'\i}guez-Fortu{\~n}o}, \citenamefont
  {Marino}, \citenamefont {Ginzburg}, \citenamefont {O’Connor}, \citenamefont
  {Mart{\'\i}nez}, \citenamefont {Wurtz},\ and\ \citenamefont
  {Zayats}}]{rodriguez2013near}%
  \BibitemOpen
  \bibfield  {author} {\bibinfo {author} {\bibfnamefont {F.~J.}\ \bibnamefont
  {Rodr{\'\i}guez-Fortu{\~n}o}}, \bibinfo {author} {\bibfnamefont
  {G.}~\bibnamefont {Marino}}, \bibinfo {author} {\bibfnamefont
  {P.}~\bibnamefont {Ginzburg}}, \bibinfo {author} {\bibfnamefont
  {D.}~\bibnamefont {O’Connor}}, \bibinfo {author} {\bibfnamefont
  {A.}~\bibnamefont {Mart{\'\i}nez}}, \bibinfo {author} {\bibfnamefont {G.~A.}\
  \bibnamefont {Wurtz}},\ and\ \bibinfo {author} {\bibfnamefont {A.~V.}\
  \bibnamefont {Zayats}},\ }\bibfield  {title} {\bibinfo {title} {Near-field
  interference for the unidirectional excitation of electromagnetic guided
  modes},\ }\href@noop {} {\bibfield  {journal} {\bibinfo  {journal} {Science}\
  }\textbf {\bibinfo {volume} {340}},\ \bibinfo {pages} {328} (\bibinfo {year}
  {2013})}\BibitemShut {NoStop}%
\bibitem [{\citenamefont {Rodr{\'\i}guez-Fortu{\~n}o}\ \emph
  {et~al.}(2014)\citenamefont {Rodr{\'\i}guez-Fortu{\~n}o}, \citenamefont
  {Puerto}, \citenamefont {Griol}, \citenamefont {Bellieres}, \citenamefont
  {Mart{\'\i}},\ and\ \citenamefont {Mart{\'\i}nez}}]{rodriguez2014sorting}%
  \BibitemOpen
  \bibfield  {author} {\bibinfo {author} {\bibfnamefont {F.~J.}\ \bibnamefont
  {Rodr{\'\i}guez-Fortu{\~n}o}}, \bibinfo {author} {\bibfnamefont
  {D.}~\bibnamefont {Puerto}}, \bibinfo {author} {\bibfnamefont
  {A.}~\bibnamefont {Griol}}, \bibinfo {author} {\bibfnamefont
  {L.}~\bibnamefont {Bellieres}}, \bibinfo {author} {\bibfnamefont
  {J.}~\bibnamefont {Mart{\'\i}}},\ and\ \bibinfo {author} {\bibfnamefont
  {A.}~\bibnamefont {Mart{\'\i}nez}},\ }\bibfield  {title} {\bibinfo {title}
  {Sorting linearly polarized photons with a single scatterer},\ }\href@noop {}
  {\bibfield  {journal} {\bibinfo  {journal} {Optics letters}\ }\textbf
  {\bibinfo {volume} {39}},\ \bibinfo {pages} {1394} (\bibinfo {year}
  {2014})}\BibitemShut {NoStop}%
\bibitem [{\citenamefont {Rodriguez-Fortuno}\ \emph {et~al.}(2014)\citenamefont
  {Rodriguez-Fortuno}, \citenamefont {Barber-Sanz}, \citenamefont {Puerto},
  \citenamefont {Griol},\ and\ \citenamefont
  {Mart{\'\i}nez}}]{rodriguez2014resolving}%
  \BibitemOpen
  \bibfield  {author} {\bibinfo {author} {\bibfnamefont {F.~J.}\ \bibnamefont
  {Rodriguez-Fortuno}}, \bibinfo {author} {\bibfnamefont {I.}~\bibnamefont
  {Barber-Sanz}}, \bibinfo {author} {\bibfnamefont {D.}~\bibnamefont {Puerto}},
  \bibinfo {author} {\bibfnamefont {A.}~\bibnamefont {Griol}},\ and\ \bibinfo
  {author} {\bibfnamefont {A.}~\bibnamefont {Mart{\'\i}nez}},\ }\bibfield
  {title} {\bibinfo {title} {Resolving light handedness with an on-chip silicon
  microdisk},\ }\href@noop {} {\bibfield  {journal} {\bibinfo  {journal} {ACS
  Photonics}\ }\textbf {\bibinfo {volume} {1}},\ \bibinfo {pages} {762}
  (\bibinfo {year} {2014})}\BibitemShut {NoStop}%
\bibitem [{\citenamefont {Mitsch}\ \emph {et~al.}(2014)\citenamefont {Mitsch},
  \citenamefont {Sayrin}, \citenamefont {Albrecht}, \citenamefont
  {Schneeweiss},\ and\ \citenamefont {Rauschenbeutel}}]{mitsch2014quantum}%
  \BibitemOpen
  \bibfield  {author} {\bibinfo {author} {\bibfnamefont {R.}~\bibnamefont
  {Mitsch}}, \bibinfo {author} {\bibfnamefont {C.}~\bibnamefont {Sayrin}},
  \bibinfo {author} {\bibfnamefont {B.}~\bibnamefont {Albrecht}}, \bibinfo
  {author} {\bibfnamefont {P.}~\bibnamefont {Schneeweiss}},\ and\ \bibinfo
  {author} {\bibfnamefont {A.}~\bibnamefont {Rauschenbeutel}},\ }\bibfield
  {title} {\bibinfo {title} {Quantum state-controlled directional spontaneous
  emission of photons into a nanophotonic waveguide},\ }\href@noop {}
  {\bibfield  {journal} {\bibinfo  {journal} {Nature communications}\ }\textbf
  {\bibinfo {volume} {5}},\ \bibinfo {pages} {1} (\bibinfo {year}
  {2014})}\BibitemShut {NoStop}%
\bibitem [{\citenamefont {Petersen}\ \emph {et~al.}(2014)\citenamefont
  {Petersen}, \citenamefont {Volz},\ and\ \citenamefont
  {Rauschenbeutel}}]{petersen2014chiral}%
  \BibitemOpen
  \bibfield  {author} {\bibinfo {author} {\bibfnamefont {J.}~\bibnamefont
  {Petersen}}, \bibinfo {author} {\bibfnamefont {J.}~\bibnamefont {Volz}},\
  and\ \bibinfo {author} {\bibfnamefont {A.}~\bibnamefont {Rauschenbeutel}},\
  }\bibfield  {title} {\bibinfo {title} {Chiral nanophotonic waveguide
  interface based on spin-orbit interaction of light},\ }\href@noop {}
  {\bibfield  {journal} {\bibinfo  {journal} {Science}\ }\textbf {\bibinfo
  {volume} {346}},\ \bibinfo {pages} {67} (\bibinfo {year} {2014})}\BibitemShut
  {NoStop}%
\bibitem [{\citenamefont {Kapitanova}\ \emph {et~al.}(2014)\citenamefont
  {Kapitanova}, \citenamefont {Ginzburg}, \citenamefont
  {Rodr{\'\i}guez-Fortu{\~n}o}, \citenamefont {Filonov}, \citenamefont
  {Voroshilov}, \citenamefont {Belov}, \citenamefont {Poddubny}, \citenamefont
  {Kivshar}, \citenamefont {Wurtz},\ and\ \citenamefont
  {Zayats}}]{kapitanova2014photonic}%
  \BibitemOpen
  \bibfield  {author} {\bibinfo {author} {\bibfnamefont {P.~V.}\ \bibnamefont
  {Kapitanova}}, \bibinfo {author} {\bibfnamefont {P.}~\bibnamefont
  {Ginzburg}}, \bibinfo {author} {\bibfnamefont {F.~J.}\ \bibnamefont
  {Rodr{\'\i}guez-Fortu{\~n}o}}, \bibinfo {author} {\bibfnamefont {D.~S.}\
  \bibnamefont {Filonov}}, \bibinfo {author} {\bibfnamefont {P.~M.}\
  \bibnamefont {Voroshilov}}, \bibinfo {author} {\bibfnamefont {P.~A.}\
  \bibnamefont {Belov}}, \bibinfo {author} {\bibfnamefont {A.~N.}\ \bibnamefont
  {Poddubny}}, \bibinfo {author} {\bibfnamefont {Y.~S.}\ \bibnamefont
  {Kivshar}}, \bibinfo {author} {\bibfnamefont {G.~A.}\ \bibnamefont {Wurtz}},\
  and\ \bibinfo {author} {\bibfnamefont {A.~V.}\ \bibnamefont {Zayats}},\
  }\bibfield  {title} {\bibinfo {title} {Photonic spin hall effect in
  hyperbolic metamaterials for polarization-controlled routing of subwavelength
  modes},\ }\href@noop {} {\bibfield  {journal} {\bibinfo  {journal} {Nature
  communications}\ }\textbf {\bibinfo {volume} {5}},\ \bibinfo {pages} {1}
  (\bibinfo {year} {2014})}\BibitemShut {NoStop}%
\bibitem [{\citenamefont {Espinosa-Soria}\ and\ \citenamefont
  {Martinez}(2016)}]{espinosa2016transverse}%
  \BibitemOpen
  \bibfield  {author} {\bibinfo {author} {\bibfnamefont {A.}~\bibnamefont
  {Espinosa-Soria}}\ and\ \bibinfo {author} {\bibfnamefont {A.}~\bibnamefont
  {Martinez}},\ }\bibfield  {title} {\bibinfo {title} {Transverse spin and
  spin-orbit coupling in silicon waveguides},\ }\href@noop {} {\bibfield
  {journal} {\bibinfo  {journal} {IEEE Photonics Technology Letters}\ }\textbf
  {\bibinfo {volume} {28}},\ \bibinfo {pages} {1561} (\bibinfo {year}
  {2016})}\BibitemShut {NoStop}%
\bibitem [{\citenamefont {Coles}\ \emph {et~al.}(2016)\citenamefont {Coles},
  \citenamefont {Price}, \citenamefont {Dixon}, \citenamefont {Royall},
  \citenamefont {Clarke}, \citenamefont {Kok}, \citenamefont {Skolnick},
  \citenamefont {Fox},\ and\ \citenamefont {Makhonin}}]{coles2016chirality}%
  \BibitemOpen
  \bibfield  {author} {\bibinfo {author} {\bibfnamefont {R.}~\bibnamefont
  {Coles}}, \bibinfo {author} {\bibfnamefont {D.}~\bibnamefont {Price}},
  \bibinfo {author} {\bibfnamefont {J.}~\bibnamefont {Dixon}}, \bibinfo
  {author} {\bibfnamefont {B.}~\bibnamefont {Royall}}, \bibinfo {author}
  {\bibfnamefont {E.}~\bibnamefont {Clarke}}, \bibinfo {author} {\bibfnamefont
  {P.}~\bibnamefont {Kok}}, \bibinfo {author} {\bibfnamefont {M.}~\bibnamefont
  {Skolnick}}, \bibinfo {author} {\bibfnamefont {A.}~\bibnamefont {Fox}},\ and\
  \bibinfo {author} {\bibfnamefont {M.}~\bibnamefont {Makhonin}},\ }\bibfield
  {title} {\bibinfo {title} {Chirality of nanophotonic waveguide with embedded
  quantum emitter for unidirectional spin transfer},\ }\href@noop {} {\bibfield
   {journal} {\bibinfo  {journal} {Nature communications}\ }\textbf {\bibinfo
  {volume} {7}},\ \bibinfo {pages} {1} (\bibinfo {year} {2016})}\BibitemShut
  {NoStop}%
\bibitem [{\citenamefont {Picardi}\ \emph {et~al.}(2017)\citenamefont
  {Picardi}, \citenamefont {Manjavacas}, \citenamefont {Zayats},\ and\
  \citenamefont {Rodr\'{\i}guez-Fortu\~no}}]{picardi2017unidirectional}%
  \BibitemOpen
  \bibfield  {author} {\bibinfo {author} {\bibfnamefont {M.~F.}\ \bibnamefont
  {Picardi}}, \bibinfo {author} {\bibfnamefont {A.}~\bibnamefont {Manjavacas}},
  \bibinfo {author} {\bibfnamefont {A.~V.}\ \bibnamefont {Zayats}},\ and\
  \bibinfo {author} {\bibfnamefont {F.~J.}\ \bibnamefont
  {Rodr\'{\i}guez-Fortu\~no}},\ }\bibfield  {title} {\bibinfo {title}
  {Unidirectional evanescent-wave coupling from circularly polarized electric
  and magnetic dipoles: An angular spectrum approach},\ }\href
  {https://doi.org/10.1103/PhysRevB.95.245416} {\bibfield  {journal} {\bibinfo
  {journal} {Phys. Rev. B}\ }\textbf {\bibinfo {volume} {95}},\ \bibinfo
  {pages} {245416} (\bibinfo {year} {2017})}\BibitemShut {NoStop}%
\bibitem [{\citenamefont {Bliokh}\ \emph {et~al.}(2014)\citenamefont {Bliokh},
  \citenamefont {Bekshaev},\ and\ \citenamefont
  {Nori}}]{bliokh2014extraordinary}%
  \BibitemOpen
  \bibfield  {author} {\bibinfo {author} {\bibfnamefont {K.~Y.}\ \bibnamefont
  {Bliokh}}, \bibinfo {author} {\bibfnamefont {A.~Y.}\ \bibnamefont
  {Bekshaev}},\ and\ \bibinfo {author} {\bibfnamefont {F.}~\bibnamefont
  {Nori}},\ }\bibfield  {title} {\bibinfo {title} {Extraordinary momentum and
  spin in evanescent waves},\ }\href@noop {} {\bibfield  {journal} {\bibinfo
  {journal} {Nature communications}\ }\textbf {\bibinfo {volume} {5}},\
  \bibinfo {pages} {1} (\bibinfo {year} {2014})}\BibitemShut {NoStop}%
\bibitem [{\citenamefont {Bliokh}\ and\ \citenamefont
  {Nori}(2015)}]{bliokh2015transverse}%
  \BibitemOpen
  \bibfield  {author} {\bibinfo {author} {\bibfnamefont {K.~Y.}\ \bibnamefont
  {Bliokh}}\ and\ \bibinfo {author} {\bibfnamefont {F.}~\bibnamefont {Nori}},\
  }\bibfield  {title} {\bibinfo {title} {Transverse and longitudinal angular
  momenta of light},\ }\href@noop {} {\bibfield  {journal} {\bibinfo  {journal}
  {Physics Reports}\ }\textbf {\bibinfo {volume} {592}},\ \bibinfo {pages} {1}
  (\bibinfo {year} {2015})}\BibitemShut {NoStop}%
\bibitem [{\citenamefont {Bliokh}\ \emph
  {et~al.}(2015{\natexlab{a}})\citenamefont {Bliokh}, \citenamefont
  {Rodr{\'\i}guez-Fortu{\~n}o}, \citenamefont {Nori},\ and\ \citenamefont
  {Zayats}}]{bliokh2015spin}%
  \BibitemOpen
  \bibfield  {author} {\bibinfo {author} {\bibfnamefont {K.~Y.}\ \bibnamefont
  {Bliokh}}, \bibinfo {author} {\bibfnamefont {F.~J.}\ \bibnamefont
  {Rodr{\'\i}guez-Fortu{\~n}o}}, \bibinfo {author} {\bibfnamefont
  {F.}~\bibnamefont {Nori}},\ and\ \bibinfo {author} {\bibfnamefont {A.~V.}\
  \bibnamefont {Zayats}},\ }\bibfield  {title} {\bibinfo {title} {Spin--orbit
  interactions of light},\ }\href@noop {} {\bibfield  {journal} {\bibinfo
  {journal} {Nature Photonics}\ }\textbf {\bibinfo {volume} {9}},\ \bibinfo
  {pages} {796} (\bibinfo {year} {2015}{\natexlab{a}})}\BibitemShut {NoStop}%
\bibitem [{\citenamefont {Bliokh}\ \emph
  {et~al.}(2015{\natexlab{b}})\citenamefont {Bliokh}, \citenamefont
  {Smirnova},\ and\ \citenamefont {Nori}}]{bliokh2015quantum}%
  \BibitemOpen
  \bibfield  {author} {\bibinfo {author} {\bibfnamefont {K.~Y.}\ \bibnamefont
  {Bliokh}}, \bibinfo {author} {\bibfnamefont {D.}~\bibnamefont {Smirnova}},\
  and\ \bibinfo {author} {\bibfnamefont {F.}~\bibnamefont {Nori}},\ }\bibfield
  {title} {\bibinfo {title} {Quantum spin hall effect of light},\ }\href@noop
  {} {\bibfield  {journal} {\bibinfo  {journal} {Science}\ }\textbf {\bibinfo
  {volume} {348}},\ \bibinfo {pages} {1448} (\bibinfo {year}
  {2015}{\natexlab{b}})}\BibitemShut {NoStop}%
\bibitem [{\citenamefont {Van~Mechelen}\ and\ \citenamefont
  {Jacob}(2016)}]{van2016universal}%
  \BibitemOpen
  \bibfield  {author} {\bibinfo {author} {\bibfnamefont {T.}~\bibnamefont
  {Van~Mechelen}}\ and\ \bibinfo {author} {\bibfnamefont {Z.}~\bibnamefont
  {Jacob}},\ }\bibfield  {title} {\bibinfo {title} {Universal spin-momentum
  locking of evanescent waves},\ }\href@noop {} {\bibfield  {journal} {\bibinfo
   {journal} {Optica}\ }\textbf {\bibinfo {volume} {3}},\ \bibinfo {pages}
  {118} (\bibinfo {year} {2016})}\BibitemShut {NoStop}%
\bibitem [{\citenamefont {Picardi}\ \emph {et~al.}(2018)\citenamefont
  {Picardi}, \citenamefont {Zayats},\ and\ \citenamefont
  {Rodr\'{\i}guez-Fortu\~no}}]{picardi2018janus}%
  \BibitemOpen
  \bibfield  {author} {\bibinfo {author} {\bibfnamefont {M.~F.}\ \bibnamefont
  {Picardi}}, \bibinfo {author} {\bibfnamefont {A.~V.}\ \bibnamefont
  {Zayats}},\ and\ \bibinfo {author} {\bibfnamefont {F.~J.}\ \bibnamefont
  {Rodr\'{\i}guez-Fortu\~no}},\ }\bibfield  {title} {\bibinfo {title} {Janus
  and huygens dipoles: Near-field directionality beyond spin-momentum
  locking},\ }\href {https://doi.org/10.1103/PhysRevLett.120.117402} {\bibfield
   {journal} {\bibinfo  {journal} {Phys. Rev. Lett.}\ }\textbf {\bibinfo
  {volume} {120}},\ \bibinfo {pages} {117402} (\bibinfo {year}
  {2018})}\BibitemShut {NoStop}%
\bibitem [{\citenamefont {Gao}\ \emph {et~al.}(2021)\citenamefont {Gao},
  \citenamefont {Zhou},\ and\ \citenamefont {Zhao}}]{gao2021continuous}%
  \BibitemOpen
  \bibfield  {author} {\bibinfo {author} {\bibfnamefont {Q.}~\bibnamefont
  {Gao}}, \bibinfo {author} {\bibfnamefont {Y.-S.}\ \bibnamefont {Zhou}},\ and\
  \bibinfo {author} {\bibfnamefont {L.-M.}\ \bibnamefont {Zhao}},\ }\bibfield
  {title} {\bibinfo {title} {Continuous photon hall effect on metal surface},\
  }\href@noop {} {\bibfield  {journal} {\bibinfo  {journal} {EPL (Europhysics
  Letters)}\ }\textbf {\bibinfo {volume} {133}},\ \bibinfo {pages} {27001}
  (\bibinfo {year} {2021})}\BibitemShut {NoStop}%
\bibitem [{\citenamefont {Luxmoore}\ \emph {et~al.}(2013)\citenamefont
  {Luxmoore}, \citenamefont {Wasley}, \citenamefont {Ramsay}, \citenamefont
  {Thijssen}, \citenamefont {Oulton}, \citenamefont {Hugues}, \citenamefont
  {Kasture}, \citenamefont {Achanta}, \citenamefont {Fox},\ and\ \citenamefont
  {Skolnick}}]{luxmoore2013interfacing}%
  \BibitemOpen
  \bibfield  {author} {\bibinfo {author} {\bibfnamefont {I.~J.}\ \bibnamefont
  {Luxmoore}}, \bibinfo {author} {\bibfnamefont {N.~A.}\ \bibnamefont
  {Wasley}}, \bibinfo {author} {\bibfnamefont {A.~J.}\ \bibnamefont {Ramsay}},
  \bibinfo {author} {\bibfnamefont {A.~C.~T.}\ \bibnamefont {Thijssen}},
  \bibinfo {author} {\bibfnamefont {R.}~\bibnamefont {Oulton}}, \bibinfo
  {author} {\bibfnamefont {M.}~\bibnamefont {Hugues}}, \bibinfo {author}
  {\bibfnamefont {S.}~\bibnamefont {Kasture}}, \bibinfo {author} {\bibfnamefont
  {V.~G.}\ \bibnamefont {Achanta}}, \bibinfo {author} {\bibfnamefont {A.~M.}\
  \bibnamefont {Fox}},\ and\ \bibinfo {author} {\bibfnamefont {M.~S.}\
  \bibnamefont {Skolnick}},\ }\bibfield  {title} {\bibinfo {title} {Interfacing
  spins in an ingaas quantum dot to a semiconductor waveguide circuit using
  emitted photons},\ }\href {https://doi.org/10.1103/PhysRevLett.110.037402}
  {\bibfield  {journal} {\bibinfo  {journal} {Phys. Rev. Lett.}\ }\textbf
  {\bibinfo {volume} {110}},\ \bibinfo {pages} {037402} (\bibinfo {year}
  {2013})}\BibitemShut {NoStop}%
\bibitem [{\citenamefont {Wang}\ \emph {et~al.}(2017)\citenamefont {Wang},
  \citenamefont {Jin}, \citenamefont {Li}, \citenamefont {Zhong}, \citenamefont
  {Liu}, \citenamefont {Kim}, \citenamefont {Jo}, \citenamefont {Rho},\ and\
  \citenamefont {Dong}}]{wang2017photonic}%
  \BibitemOpen
  \bibfield  {author} {\bibinfo {author} {\bibfnamefont {Y.-H.}\ \bibnamefont
  {Wang}}, \bibinfo {author} {\bibfnamefont {R.-C.}\ \bibnamefont {Jin}},
  \bibinfo {author} {\bibfnamefont {J.-Q.}\ \bibnamefont {Li}}, \bibinfo
  {author} {\bibfnamefont {F.}~\bibnamefont {Zhong}}, \bibinfo {author}
  {\bibfnamefont {H.}~\bibnamefont {Liu}}, \bibinfo {author} {\bibfnamefont
  {I.}~\bibnamefont {Kim}}, \bibinfo {author} {\bibfnamefont {Y.}~\bibnamefont
  {Jo}}, \bibinfo {author} {\bibfnamefont {J.}~\bibnamefont {Rho}},\ and\
  \bibinfo {author} {\bibfnamefont {Z.-G.}\ \bibnamefont {Dong}},\ }\bibfield
  {title} {\bibinfo {title} {Photonic spin hall effect by the spin-orbit
  interaction in a metasurface with elliptical nano-structures},\ }\href@noop
  {} {\bibfield  {journal} {\bibinfo  {journal} {Applied Physics Letters}\
  }\textbf {\bibinfo {volume} {110}},\ \bibinfo {pages} {101908} (\bibinfo
  {year} {2017})}\BibitemShut {NoStop}%
\bibitem [{\citenamefont {Zhao}\ and\ \citenamefont
  {Zhou}(2021)}]{zhao2021tunable}%
  \BibitemOpen
  \bibfield  {author} {\bibinfo {author} {\bibfnamefont {L.-M.}\ \bibnamefont
  {Zhao}}\ and\ \bibinfo {author} {\bibfnamefont {Y.-S.}\ \bibnamefont
  {Zhou}},\ }\bibfield  {title} {\bibinfo {title} {Tunable multichannel
  photonic spin hall effect in metal-dielectric-metal waveguide},\ }\href@noop
  {} {\bibfield  {journal} {\bibinfo  {journal} {Scientific Reports}\ }\textbf
  {\bibinfo {volume} {11}},\ \bibinfo {pages} {1} (\bibinfo {year}
  {2021})}\BibitemShut {NoStop}%
\end{thebibliography}
\end{document}